\documentclass[manuscript]{copernicus}

\begin{document}
\nolinenumbers
\title{Viscoelastic mechanics of tidally induced lake drainage in the Amery grounding zone}

\Author[1][hanwen.zhang@earth.ox.ac.uk]{Hanwen}{Zhang} 
\Author[1]{Richard}{Katz}
\Author[1]{Laura}{Stevens}

\affil[1]{Department of Earth Sciences, University of Oxford, South Parks Road, Oxford, OX1 3AN, UK}




\runningtitle{H.Zhang et al.}
\runningauthor{Viscoelastic mechanics of tidally induced lake drainage in the Amery grounding zone}

\received{}
\pubdiscuss{} 
\revised{}
\accepted{}
\published{}


\firstpage{1}

\renewcommand{\figureautorefname}{Fig.}
\renewcommand{\equationautorefname}{Eq.}
\renewcommand{\sectionautorefname}{Sect.}

\maketitle

\begin{abstract}
Drainage of supraglacial lakes to the ice-sheet bed can occur when a hydrofracture propagates downward, driven by the weight of the water in the lake. For supraglacial lakes in the grounding zones of Antarctic glaciers, the mechanics of drainage is complicated by their proximity to the grounding line. Recently, a series of supraglacial lake-drainage events through hydrofractures was observed in the Amery Ice Shelf grounding zone, East Antarctica. The lake depth at drainage varied considerably between events, raising questions about the mechanisms that induce hydrofracture, even at low lake depths. Here we use a modelling approach to investigate the contribution of tidally driven flexure to hydrofracture propagation. We model the viscoelastic response of a marine ice sheet to tides, the stresses that are induced, and the contribution of tidal stresses to hydrofracture propagation. Our results show that ocean tides and lake-water pressure together control supraglacial lake drainage through hydrofractures in the grounding zone. We give a model-based criterion that predicts supraglacial lake drainage as a function of daily maximum tidal amplitude and lake depth. Our model-based criterion agrees with remotely sensed data, indicating the importance of tidal flexure to processes associated with hydrofracturing such as supraglacial lake drainage, rifting and calving.
\end{abstract}


\introduction  
Atmospheric warming is driving increasing meltwater production on ice-sheet and ice-shelf surfaces \citep{trusel2015divergent}. In the melt season, the meltwater ponds in topographic lows and forms supraglacial lakes. Lakes drain either slowly, through surface drainage channels \citep{banwell2019direct}, or rapidly, through hydrofractures \citep{das2008fracture}. Lake drainages through hydrofractures can potentially impact ice-sheet mass balance in various ways. For grounded ice sheets, hydrofracture efficiently transports surface meltwater to the subglacial hydrological system. This reduces bed friction and thus modulates ice-flow velocity and flux \citep{das2008fracture,doyle2013ice,tedesco2013ice,stevens2015greenland,dunmire2020observations}. At ice shelves, hydrofractures can initiate or promote rifts. When propagating through ice shelves, rifts can destabilise them by triggering iceberg calving and ice-shelf collapse \citep{scambos2000link,glasser2008structural,banwell2013breakup,banwell2019direct,warner2021rapid,lipovsky2020ice}, which leads to a loss of buttressing and increased ice-sheet mass loss.

In East Antarctica, satellite imagery suggests that supraglacial lakes often cluster in the grounding zone, particularly at low elevations and bedslopes. Many of these lakes are connected to surface drainage systems or located in regions that are vulnerable to hydrofracturing \citep{stokes2019widespread}. In the grounding zone, besides lake--water pressure, tensile stress due to tidal flexure can promote hydrofracturing. Meanwhile, advection of damaged ice produced in the grounding zone could destabilise the downstream ice shelf \citep{borstad2012damage}. Thus it is important to understand how hydrofractures are initiated and promoted in the grounding zone.

Recently, \cite{trusel2022repeated} reported a series of repeated drainage events of a supraglacial lake at the grounding line (GL) of the Amery Ice Shelf, East Antarctica. Interestingly, these drainage events did not occur past a threshold in lake volume, but rather tended to coincide with times of high daily tidal amplitude. These observations raise the question: \textit{how do tides near the GL contribute to lake drainage through hydrofracturing}? \citet{trusel2022repeated} hypothesised that near the GL, the drainage events are promoted by tensile stress due to tidal flexure. To test this hypothesis and further explore the dependence of lake drainage on ocean tides and lake depth, we present a numerical model that accounts for the tidal contribution to hydrofracturing near the Amery Ice Shelf GL.

As an important component of the marine ice sheet, the GL serves as an internal boundary connecting the grounded ice sheet and floating ice shelf. Variations in its position control the total ice flux from inland to the sea. In Antarctica, the GL response to diurnal ocean tides has been documented by various observations. On the Rutford Ice Stream, West Antarctica, ice flow is modulated by semi-diurnal tides, with tidal effects on the ice-flow rate propagating tens of kilometres upstream \citep{gudmundsson2006fortnightly,murray2007ice,minchew2017tidally}. At the Amery Ice Shelf, kilometre-scale tidal GL migration with seawater intrusion has been observed from differential radar interferometry \citep{chen2023grounding}. The observed grounding zone is much larger than predicted from hydrostatic equilibrium, raising questions about whether the observed tidal flexure within the grounding zone is associated with stresses that contribute to hydrofracturing.

Ice-shelf flexure at the grounding line can be modelled using thin-plate theory with an elastic \citep{vaughan1995tidal,sayag2011elastic,wagner2016role,warburton2020tidal} or a viscoelastic constitutive relationship \citep{reeh2003tidal,gudmundsson2007tides}. In these models, the GL is treated as a peeling front or as the clamped end of the ice shelf. Beam models capture the large-scale flexure of ice shelves, and neglect the membrane stress that is governed by sliding.

Various studies have used a vertically integrated theory to investigate the dependence of steady-state grounding-line position on ice thickness, sliding laws, buttressing effects and bed topography \citep{schoof2007ice,schoof2007marine,katz2010stability, schoof2012marine,tsai2015marine,pegler2018marine,haseloff2022effects,sergienko2023stable}. Besides depth-integrated models, full-Stokes models have also been used to study the migration of grounding lines on both longer timescales
\citep{nowicki2008conditions, durand2009marine,favier2012three,gudmundsson2012stability,cheng2020full} and tidal timescales \citep{gudmundsson2011ice,rosier2014insights,rosier2015temporal,rosier2020exploring}. In these models,  the ice-sheet--bed contact problem has been implemented with boundary conditions. A more recent development by \citet{stubblefield2021variational} and \citet{de_diego_farrell_hewitt_2022} has incorporated the contact boundary conditions into variational inequalities. This formulation enabled the representation of the contact condition within a variational framework that was implemented in Finite Element method. In this study, we adopt the framework by \citet{stubblefield2021variational} to study the GL dynamics at the Amery Ice Shelf.

Since the semi-diurnal tidal period at the Amery Ice Shelf (approximately $12$ hours) is close to the Maxwell time of ice (approximately $9$ hours in our estimation), we use a viscoelastic constitutive relationship to model the tidal flexure. In particular, we extend the framework for a marine ice sheet with viscous ice flow by \citet{stubblefield2021variational} to an upper-convected Maxwell model to capture the stress and GL migration. We predict the tensile stress at the GL with daily maximum tidal amplitudes. Using Linear Elastic Fracture Mechanics (LEFM) analysis, we estimate the contributions from tidal stress and lake-water supply to quasi-static hydrofracturing. This enables a model-based criterion for supraglacial lake drainage in terms of tidal amplitude and lake-water depth, which we compare to the lake-drainage time series presented by \citet{trusel2022repeated}. The results support the hypothesis that at the Amery Ice Shelf GL, supraglacial lake drainage is controlled by both lake depth and tidal amplitude. The relative importance of the two factors can be estimated using the model-based criterion.

The paper is organised as follows. In \autoref{sec:method}, we show an analysis of the ice-flow patterns near the lake, the viscoelastic marine ice-sheet model, and the corresponding numerical implementation. In \autoref{sec:results}, we demonstrate the viscoelastic tidal response of a marine ice sheet, and then use LEFM analysis to construct a drainage criterion and compare it with observations. In \autoref{sec:discussion}, we present an analysis of the model's sensitivity to ice Maxwell time and bedslope angle, and summarise model limitations.

\section{Method}\label{sec:method}
In this section, using remotely-sensed ice-surface velocity fields, we first present an estimation of the local strain-rate and stress fields near the lake studied by \citet{trusel2022repeated}. Our estimation indicates the lake region is extension-dominated with negligible shear stress. The background stress is insufficient to produce hydrofracturing, indicating the importance of tidal stress. Therefore, we adopt the viscous, marine ice-sheet model by \citet{stubblefield2021variational} and incorporate a viscoelastic rheological formulation. We demonstrate the model set-up, how lake drainage events are predicted, and how the observational data is processed to enable comparison with models.

\subsection{Data extraction and model selection}
\begin{figure*}[t]
    \includegraphics[width=16cm]{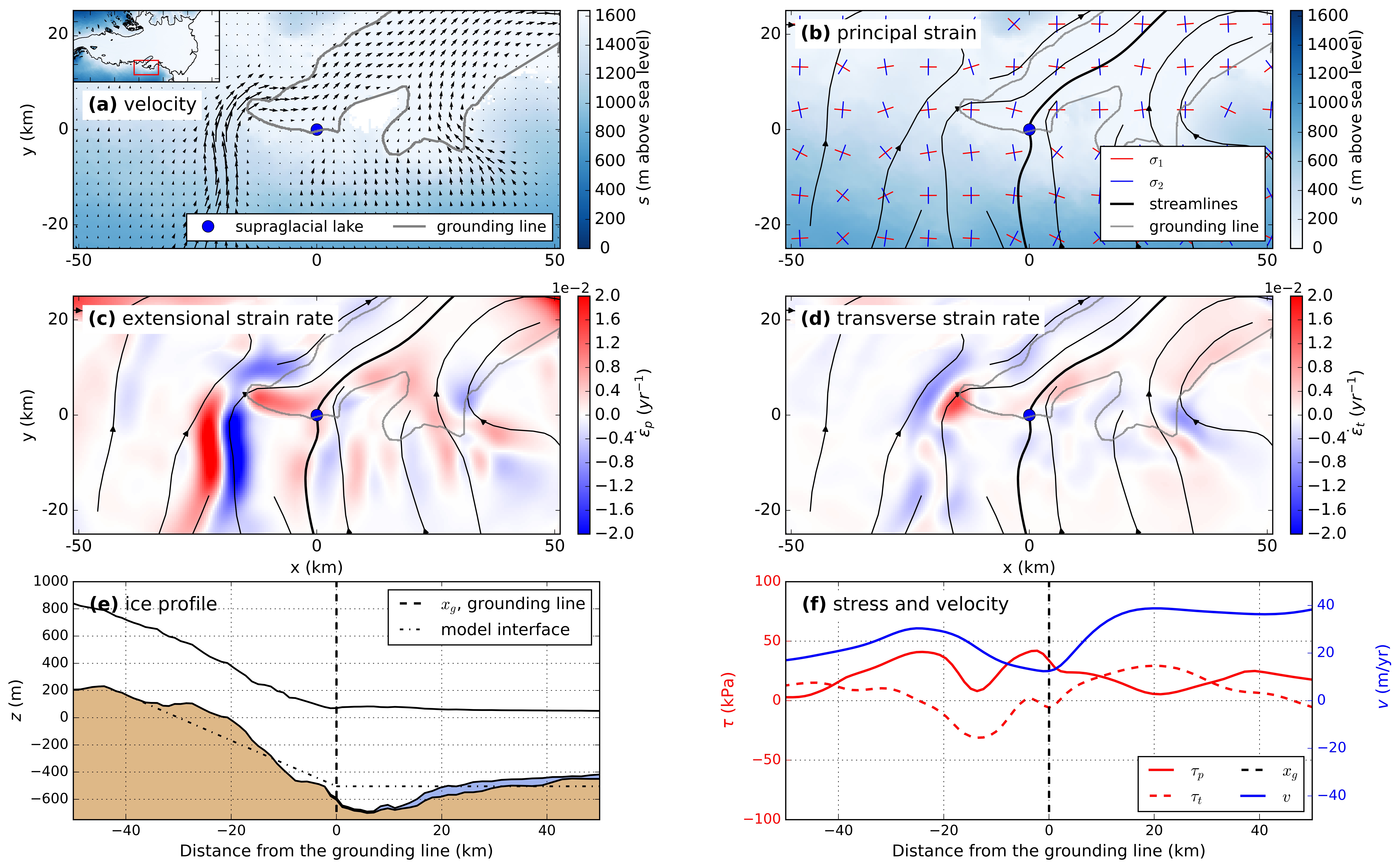}
    \caption{Ice surface velocity, strain rate and stress in the region of the supraglacial lake. \textbf{(a)} Velocity field near the grounding line (grey), where the supraglacial lake is denoted with the blue dot. The color represents the ice-sheet surface elevation above sea level. The map inset on the top left corner shows the full Amery Ice Shelf topography, with the plotted region outlined with a red box; \textbf{(b)} principal strain rate and streamlines. The streamline that crosses the lake is marked with the bold line; \textbf{(c)} along-flow strain rate; \textbf{(d)} transverse strain rate; \textbf{(e)} local ice-sheet geometry and bed topography; \textbf{(f)} For the streamline that crosses the lake, along-flow deviatoric extension $\sigma_p$ (solid red line) and shear stress $\sigma_t$ (dashed red line), and speed $v$ (blue). Note that $x=x_g=0$ is the position of the supraglacial lake as well as the grounding line.}
    \label{fig:amery}
\end{figure*}

Figure \ref{fig:amery}(a) shows the ice-surface velocity field $\boldsymbol{v}$ \citep{rignot2011ice,Rignot2017velocity,mouginot2012mapping,mouginot2017comprehensive} near the lake, according to which we calculate the streamlines and the strain rate ${\dot{\boldsymbol{\varepsilon}}}$ (\autoref{fig:amery}b). The lake is located on the grounding line extracted from \citet{Rignot2016groundingline,rignot2011antarctic,rignot2014widespread,li2015grounding}. Following \citet{wearing2017flow}, we denote the direction of velocity $\hat{\boldsymbol{v}}$, and the transverse direction $\hat{\boldsymbol{t}}$. The along-flow strain rate can be defined as 
\begin{equation}
    \dot{\varepsilon}_p = \hat{\boldsymbol{v}}\cdot {\boldsymbol{\dot{\varepsilon}}}\cdot\hat{\boldsymbol{v}},
\end{equation}
and the transverse strain rate as 
\begin{equation}
    \dot{\varepsilon}_t = \hat{\boldsymbol{v}}\cdot {\boldsymbol{\dot{\varepsilon}}}\cdot\hat{\boldsymbol{t}}.
\end{equation}

Figure \ref{fig:amery}(c,d) shows $\dot{\varepsilon}_p$ and $\dot{\varepsilon}_t$ near the lake. We are particularly interested in $\dot{\varepsilon}_p$ and $\dot{\varepsilon}_t$ along the streamline that crosses the lake. Assuming that the $z$-components of ice deviatoric stress $\boldsymbol \tau$ are zero on the surface ($\tau_{xz}=\tau_{yz}=\tau_{zz}=0$), we calculate the viscosity $\eta$ using Glen's flow law \citep{glen1955creep}. The deviatoric stress along ($\tau_p$) and transverse ($\tau_t$) to the streamline are $\tau_{p} = 2\eta \dot{\varepsilon}_{p},~\tau_{t} = 2\eta\dot{\varepsilon}_{t}$, as shown in \autoref{fig:amery}(f). The magnitude of $\tau_t$ and $\tau_p$ are less than $40$ kPa along the streamline. Specifically, at the lake, there is negligible background shear stress and an extensional stress with a magnitude of about $30$ kPa in the flow direction. Therefore, for simplicity, we neglect the background stresses, focus on the the streamline and adopt a 2-D flow line model that accounts for tidally induced extension along the flow direction.

We use BedMachine Antarctica v2 to obtain the basal topography and ice geometry along the streamline (\autoref{fig:amery}e) \citep{morlighem2017bedmachine, morlighem2020deep}. The surface gradient is relatively uniform upstream of the GL. The subglacial cavities downstream of the grounding zone are more than $20$ m wide. Thus, we assume that the ice shelf downstream of the GL doesn't contact the bedrock, and the water pressure on the ice--ocean interface is hydrostatic. In the computation, we use a linear bedrock topography (\autoref{fig:amery}e) with a slope angle extracted from reality. This simplification enables us to further explore the sensitivity of GL dynamics to bedslope angle. In \ref{apdx:real_bed} we present a set of model results with real bed topography for comparison. While the magnitude of tidal stress is modified quantitatively in the real-bed-topography case, tidal stress remains the main contribution to the extension on the GL, compared with the background extensional and shear stress.

\subsection{Model domain}\label{subsec:model_domain}
\begin{figure}[t]
    \includegraphics[width=10cm]{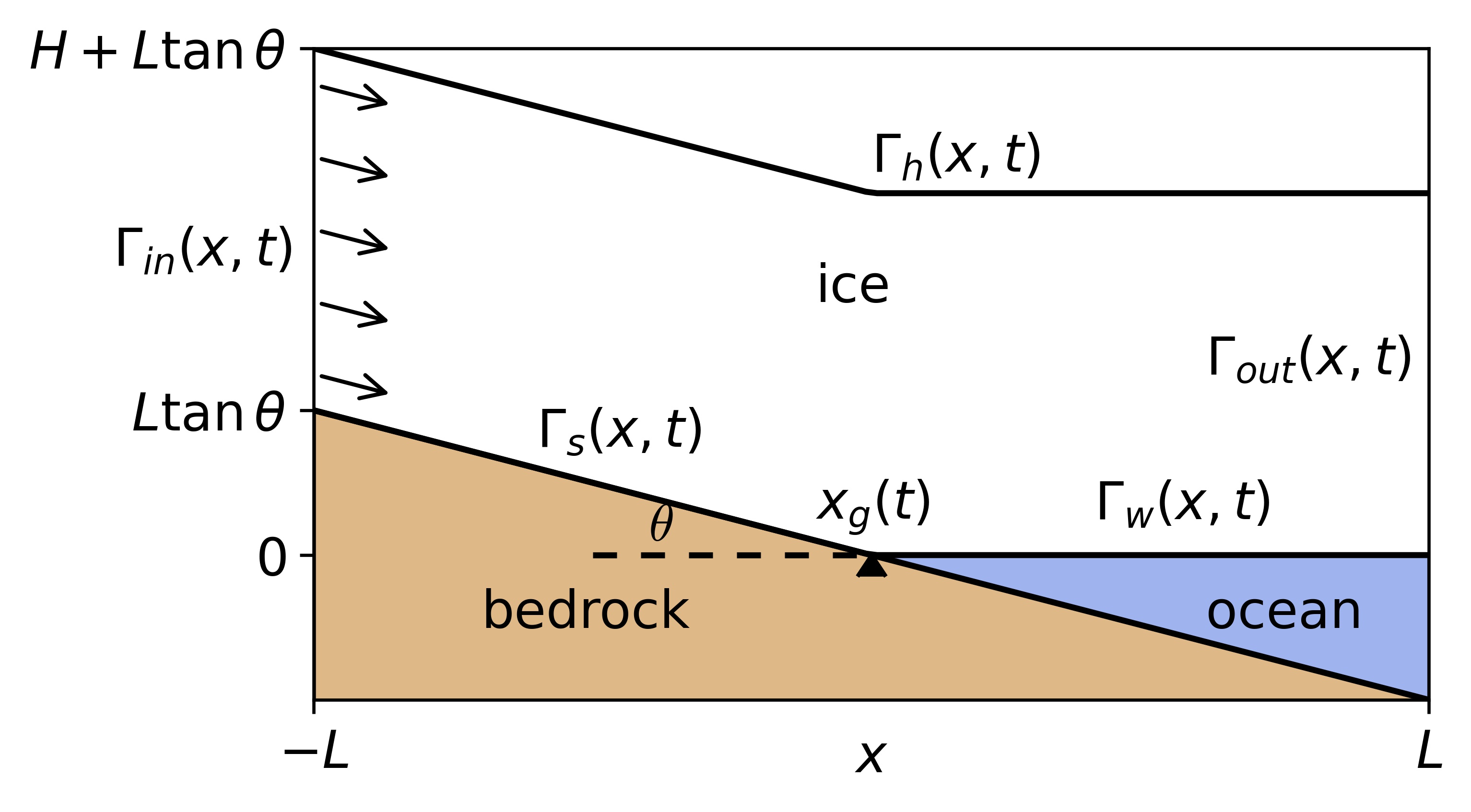}
    \caption{Schematic showing the model domain of a marine ice sheet system.}
    \label{fig:schematic}
\end{figure}

Figure \ref{fig:schematic} shows a schematic of the computational domain. We consider a segment of marine ice sheet with length $2L$ and thickness $H(x,t)$ in a Cartesian coordinate system with position vector $\boldsymbol x =\left(x,z\right)$, where $z$ increases upward. The inflow and outflow boundaries are denoted $\Gamma_{in}$ and $\Gamma_{out}$. The top surface is denoted $\Gamma_h$. The bottom is divided into two parts, according to whether the ice is in contact with the bedrock or the ocean. The ice--bedrock interface $\Gamma_{s}$ is where ice is in contact with the bedrock at a height $b\left(x\right)$. As a simplification, we assume that the bedrock has a uniform slope $\theta$. The ice--ocean interface $\Gamma_{w}$ is where ice is detached from the bedrock. The two boundaries $\Gamma_{s}$ and $\Gamma_{w}$ meet at the GL whose horizontal position, denoted $x_g\left(t\right)$, migrates with time $t$. The origin of the coordinate system is set at the middle of the domain on the ice--bedrock interface (at the position of the GL shown in the schematic).

\subsection{Governing equations}\label{subsec:gov_eq}
The governing equations for momentum and mass conservation are
\begin{align}
\boldsymbol{\nabla}\cdot{\boldsymbol{\sigma}}+\rho_i \boldsymbol{g}=\boldsymbol{0},\\
\boldsymbol{\nabla}\cdot{\boldsymbol{u}}=0,
\end{align}
where $\boldsymbol{\sigma}$ is the total Cauchy stress tensor, $\rho_i$ is ice density, $\boldsymbol{g}$ is gravity, and $\boldsymbol{u}$ is the ice velocity field. The stress $\boldsymbol{\sigma}$ can be decomposed as an isotropic part and deviatoric part, $\boldsymbol{\sigma}=-p\boldsymbol I+ \boldsymbol{\tau},$ where $p$ and $\boldsymbol{\tau}$ represent the pressure and deviatoric stress, respectively. Here $\boldsymbol{I}$ is the unit tensor.

To model viscoelasticity, we adopt the upper-convected Maxwell formulation for the deviatoric stress $\boldsymbol{\tau}$. The constitutive relationship is
\begin{equation}
    \boldsymbol\tau+\lambda \overset{\nabla}{\boldsymbol\tau}=2 \eta\dot{\boldsymbol{\varepsilon}},
\end{equation}
where the Maxwell time, $\lambda=\eta/\mu$, is the ratio of ice viscosity to shear modulus. The strain rate is denoted $\dot{\boldsymbol{\varepsilon}}$. The upper-convected time derivative (Oldroyd rate) $\overset{\nabla}{\boldsymbol\tau}$ measures the temporal variation of $\boldsymbol{\tau}$ including the effect of rigid body rotation,
\begin{equation}\label{eq:rheology}
    \overset{\nabla}{\boldsymbol\tau}=\partial_{t} \boldsymbol{\tau}+\boldsymbol{u} \cdot \nabla \boldsymbol{\tau}-(\nabla \boldsymbol{u})^{T} \cdot \boldsymbol{\tau}-\boldsymbol{\tau} \cdot \nabla \boldsymbol{u},
\end{equation}
where $\left(\cdot\right)^T$ represents tensor transpose.

We assume a constant shear modulus $\mu$ and non-Newtonian viscosity $\eta$ that is governed by Glen's flow law with regularisation
\begin{equation}\label{eq:viscosity}
    \eta =\frac{1}{2} B\left(|\boldsymbol{\dot{\varepsilon}}|^2+\delta_{\nu}\right)^{-{(n-1)}/{2n}},
\end{equation}
where $B=2^{(n-1) / 2 n} A_0^{-1 / n}$ is determined by the two flow law parameters $A_0$ and $n$, and $|\boldsymbol{\dot{\varepsilon}}| =\sqrt{\boldsymbol{\dot{\varepsilon}}:\boldsymbol{\dot{\varepsilon}}}$ is the Frobenius norm of the strain rate. The numerical parameter $\delta_{\nu}$ is used to prevent infinite viscosity at vanishing strain rate \citep{jouvet2011analysis, helanow2018stabilized, stubblefield2021variational}. The value of $\delta_{\nu}$ sets an upper limit on the viscosity and, therefore, also on the Maxwell time. In our reference parameter set, used below, $\eta\le 4.3\times 10^{13}~\text{Pa}~\text{s}$ and $\lambda\le 40$ hr. When $\delta_{\nu}=0$, Eq. \eqref{eq:viscosity} reduces to the classical form of Glen's flow law.

\subsection{Boundary conditions}
Neglecting atmospheric pressure and other surface loading, the top boundary is assumed to be stress-free. Its elevation $h$ is governed by the kinematic condition
\begin{equation}\label{eq:kinematic_top}
\frac{\partial h}{\partial t}(x, t)= \left[\left(\frac{\partial h}{\partial x}\right)^2+1\right]^{1/2} \boldsymbol{u}\cdot \boldsymbol{n}\qquad\text{on }\Gamma_h,
\end{equation}
where $\boldsymbol n$ is the outward-pointing unit normal vector.

On the inflow boundary $\Gamma_{in}$, we impose a uniform horizontal inflow rate $u_0$ and zero shear stress
\begin{equation}\label{eq:subglacial_bc_left}
\left\{\begin{array}{c}
\boldsymbol{u}\cdot \boldsymbol{n} = u_0,\\
    \boldsymbol t\cdot \boldsymbol \sigma\cdot \boldsymbol n = 0,
\end{array}\right.\qquad \text{on}~\Gamma_{in},
\end{equation}
where $\boldsymbol{t}$ is the tangential unit vector. The inflow velocity $u_0$ is set to be the satellite-derived surface velocity, $9$~$\text{m}\,\text{y}^{-1}$ \citep{Rignot2016MEaSUREs}.
On the outflow boundary, we impose the ice-overburden pressure
\begin{equation}\label{eq:subglacial_bc_right}
    \boldsymbol \sigma \cdot \boldsymbol n = -\rho_i g \left(h-z\right)\boldsymbol n,\qquad \text{on}~\Gamma_{out},
\end{equation}
which means that at the downstream boundary, the ice shelf floats at hydrostatic equilibrium, without bending stress.

Similar to Eq. \eqref{eq:kinematic_top}, the bottom profile $s\left(x,t\right)$ is governed by the kinematic equation
\begin{equation}\label{eq:kinematic_bottom}
\frac{\partial s}{\partial t}(x, t)= - \left[\left(\frac{\partial s}{\partial x}\right)^2+1\right]^{1/2} \boldsymbol{u}\cdot \boldsymbol{n},\qquad \text{on }\Gamma_w.
\end{equation}
The stress on the bottom boundary depends on the local contact condition. To introduce the boundary conditions related to the contact problem, we consider hydrostatic water pressure $p_w$ on the ice--ocean interface, defined as
\begin{equation}\label{eq:hydrology}
    p_w = \rho_w g\left[h_w\left(t\right)-s^{*}\right], \qquad \text{on }\Gamma_w,
\end{equation}
where $\rho_w$ is water density, $g$ is gravitational acceleration, $h_w$ is the sea level and $s^{*}$ is the approximated bottom boundary that will be introduced in \autoref{subsec:numerical}.

The sea level $h_w$ is a superposition of a steady state $h_0$ and a sinusoidal function of time, representing ocean tides with amplitude $A$ and frequency $f$,
\begin{equation}\label{eq:tide}
    h_w\left(t\right)=h_0 + A \sin\left(2\pi f t\right).
\end{equation}

On the ice--ocean interface, the hydrostatic pressure $p_w$ is imposed as the traction
\begin{equation}
    \boldsymbol\sigma \cdot \boldsymbol n = -p_w\boldsymbol n, \qquad\text{on }\Gamma_w.
\end{equation}

On the ice--bedrock interface, ice can be either attached or detached from the bed. In the normal direction, the contact condition is established by the following boundary conditions
\begin{equation}\label{eq:contact_bc}
\left\{\begin{array}{c}
\sigma_n \geqslant p_w, \\
u_n\le 0, \\
\left(\sigma_n-p_w\right)u_n=0,
\end{array}\right.\qquad\text{on }\Gamma_s,
\end{equation}
where $\sigma_n$ is the normal component of traction. The contact condition is implemented using a penalty term shown in \autoref{subsec:numerical}, originally proposed by \citet{stubblefield2021variational}.

In the tangential direction, ice sliding is resisted by friction that is governed by a Weertman-type sliding law \citep{weertman1957sliding}
\begin{equation}\label{eq:basal_friction}
    \boldsymbol t\cdot \boldsymbol \sigma\cdot \boldsymbol n = -C\left[\left(\boldsymbol u\cdot \boldsymbol t\right)^2+\delta_s\right]^{-\frac{n-1}{2n}}\boldsymbol u\cdot \boldsymbol t \qquad\text{on }\Gamma_s,
\end{equation}
where $C$ is the friction coefficient and $\delta_s$ is a numerical factor preventing singularity. In the computation, we choose $C$ such that the surface velocity at the GL matches the inflow speed $u_0$. This choice gives a relatively low surface velocity gradient, which agrees with satellite observations at the lake region \citep{Rignot2016MEaSUREs}.

\subsection{Numerical Implementation}\label{subsec:numerical}
When implementing the hydrostatic water pressure on the ice--ocean interface, for numerical stability, rather than the bottom elevation from the previous time step, $s^{*}$ is an approximation to the current step elevation \citep{durand2009marine,stubblefield2021variational}, defined as
\begin{equation}
s_*(x, t)=s(x, t-\Delta t)-u_n(x, s, t)\Delta t,
\end{equation}
where $\Delta t$ is the numerical time step, $s(x, t-\Delta t)$ is the bottom profile at the previous time step, and $u_n$ is the normal velocity on the bottom boundary. 

In the variational formulation, the contact condition Eq. \eqref{eq:contact_bc} is accounted for by adding a line integral along the ice--bedrock interface as a penalty term
\begin{equation}\label{eq:penalty}
    \frac{1}{\varepsilon}\int_{\Gamma_s} \frac{1}{2}\left(\boldsymbol{u}\cdot\boldsymbol{n}+\left|\boldsymbol{u}\cdot\boldsymbol{n}\right|\right)\boldsymbol{v}\cdot\boldsymbol{n}~\mathrm{d}\Gamma_{s},
\end{equation}
where $\varepsilon$ is the penalty parameter and $\boldsymbol{v}$ is the test function corresponding to the velocity field $\boldsymbol{u}$. As $\varepsilon\rightarrow 0$, the solution to the variational formulation weakly converges to the solution governed by the contact condition Eq. \eqref{eq:contact_bc} \citep{kikuchi1988contact}.

The variational formulation is implemented using the finite-element library FEniCS \citep{logg2010dolfin,logg2012automated,LangtangenLogg2017}. A mixed finite element is used to solve for a combined field $\left(\boldsymbol u,~p,~\boldsymbol \tau\right)$. We use triangular elements in which the pressure varies linearly and the velocity and deviatoric stress vary quadratically. In \ref{apdx:convergence}, we report convergence tests showing the results are mesh-independent and, in the limit of no elastic deformation ($\mu\rightarrow\infty$), converge to the viscous solutions by \citet{stubblefield2021variational}. For further details about the variational formulation and its numerical implementation, the reader is referred to \cite{stubblefield2021variational}.

%
%

\begin{table*}[t]
\centering
\caption{Parameters used in numerical model and their reference values.}
\begin{tabular}{l c r}
    \tophline
    Physical property
      & Notation
      & Value \\ \middlehline
    Density of water & $\rho_w$ & $1000$ $\text{kg}~ \text{m}^{-3}$\\
    Density of ice & $\rho_i$ & $917$ $\text{kg}~\text{m}^{-3}$\\
    Length of the domain & $L$ & $20$ $\text{km}$\\
    Ice thickness & $H$ & $500$ $\text{m}$\\
    Bedslope angle & $\theta$ & 0.02\\
    Glen's Law exponent & $n$ & $3$\\
    Viscosity coefficient & $A_0$ & $3.1689\times 10^{-24}$ $\text{Pa}^{-n}~ \text{s}^{-1}$\\
    Characteristic (inflow) velocity & $u_0$ & $9\ \text{m}~\text{y}^{-1}$\\
    Friction coefficient & $C$ & $1.0\times10^7$ $\text{Pa}^{1/n}~ \text{m}^{-1}$\\
    Shear modulus & $\mu$ & $0.30\times 10^{9}$ $\text{Pa}$\\
    Viscosity regularisation parameter & $\delta_{\nu}$ & $10^{-18}$\\
    Friction regularisation parameter & $\delta_{s}$ & $10^{-15}$\\
    Penalty parameter & $\varepsilon$ & $10^{-13}$\\
    Tidal amplitude & $A$ & 1~\text{m}\\ \bottomhline
\end{tabular}
\label{tab:constants}
\belowtable{} 
\end{table*}

\subsection{Lake basin bathymetry}
For the drainage events in \citet{trusel2022repeated}, we calculate lake-basin depth from a water-free, 2-m WorldView-1 DEM. We choose the elevation of the flat basin, excluding craters and hydrofractures, as the deepest point of the lake. The lake, located at 70.59 \textdegree{}S, 72.53 \textdegree{}E, has a depth of about $10$~m and a width of about $1$~km in the direction normal to the GL. Using a shoreline-extraction method \citep{moussavi2016derivation,trusel2022repeated}, water depth is calculated as the difference between the median shoreline elevation and the lake-basin elevation.

\section{Results}\label{sec:results}
First, we show a reference case representing the tidal response of the marine ice sheet near the lake reported in \cite{trusel2022repeated}. The lake, whose depth ($\le10$ m) is much smaller than ice thickness ($\approx 500$ m), is assumed to have no effects on GL dynamics and thus not included here. We consider a $20$-km long, $500$-m thick ice sheet sliding on  bedrock with constant bedslope angle $\theta=0.02$. The ice thickness and bedslope angle, representing the local topography, are adopted from \cite{morlighem2017bedmachine}. Initially, the grounded ice sheet and floating ice shelf are both $10$ km long. In the model of elasticity, we use Young's modulus $E=0.88$~GPa and Poisson's ratio $\nu=0.41$, as suggested by \citet{vaughan1995tidal} for tidal flexure problems. A list of of parameters and their references values is provided in \autoref{tab:constants}.

\subsection{Tidally-induced grounding line migration and stress}
\label{sec:ref_case_stress}
\begin{figure*}[t]
    \includegraphics[width=16cm]{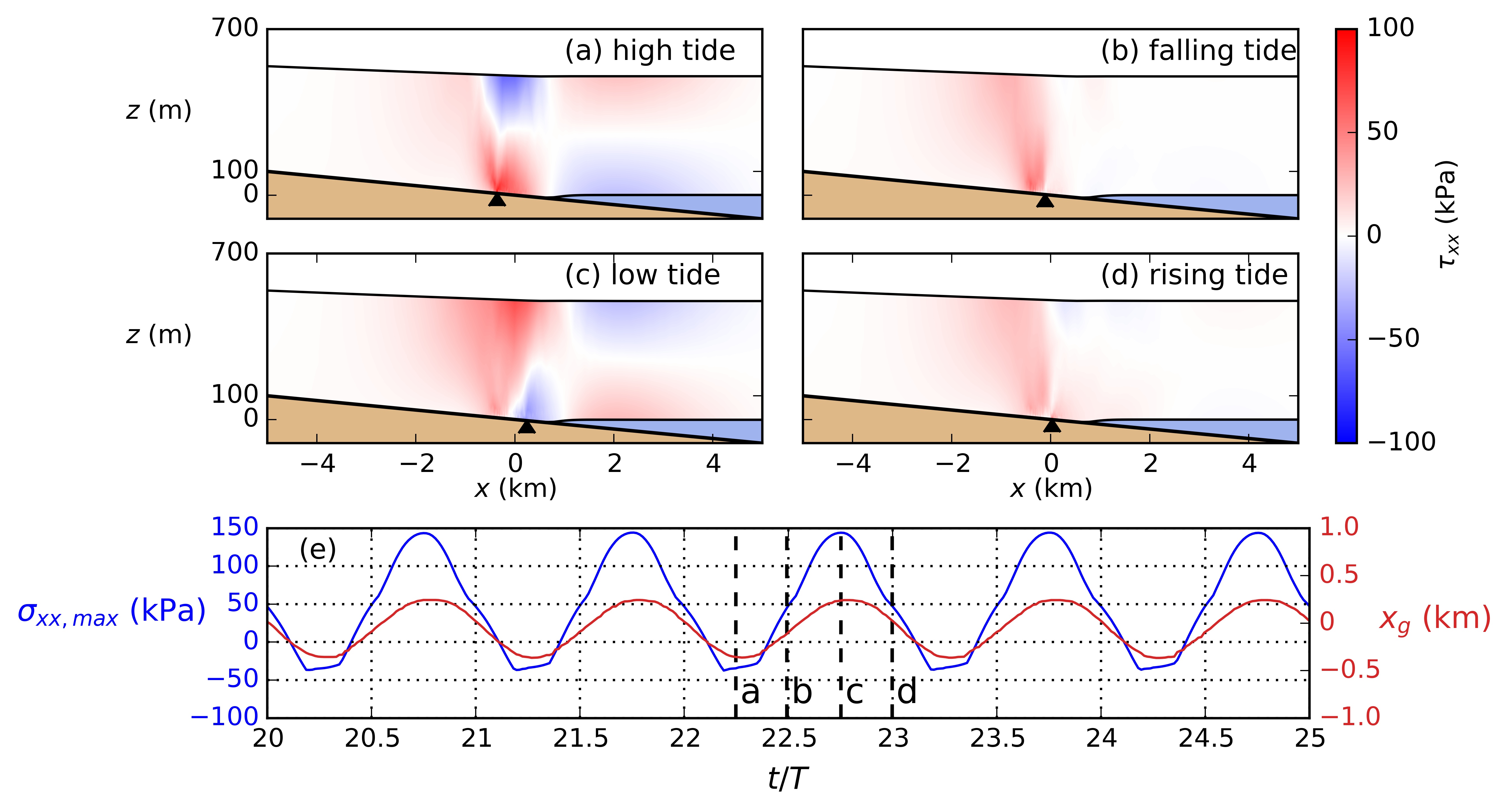}
    \caption{Tidal response of a marine ice sheet at different tidal phases. \textbf{(a)--(d)} Deviatoric tensile stress $\tau_{xx}$ in one tidal period. \textbf{(e)} The maximum tensile stress $\sigma_{xx,max}$ (blue) on the top boundary within the lake region ($\bar{x_g}-0.5~\text{km}\le x \le \bar{x_g}+ 0.5~\text{km}$) and the GL position $x_g$ (red) versus time (scaled by the tidal period $T$) with positive values representing downstream migration. Vertical dashed lines show the time of panels (a)-(d). }
    \label{fig:ref_case}
\end{figure*}

Following \cite{stubblefield2021variational}, we first construct the mesh with piecewise linear bottom profile $s\left(x\right)$ and surface profile $h\left(x\right)$
\begin{align}    
s\left(x\right)&=\mathrm{max}\left(b\left(x\right),0\right),\\
h\left(x\right)&=s\left(x\right)+H,
\end{align}
which are shown in \autoref{fig:schematic}. We evolve this initial profile with no tides for $40$~yr, by which time the ice-flow geometry (i.e., $s\left(x\right)$, $h\left(x\right)$ and GL $x_g$) reaches a steady state. This provides a steady profile of the marine ice sheet for use as an initial condition for simulations with tides.

The important dynamics consist of the tidally-modulated GL migration and corresponding stress. The GL position $x_g$ is shown in \autoref{fig:ref_case}e. Whereas \citet{stubblefield2021variational} find double GLs at low tides with a relatively small bedslope angle $\theta=2.5\times 10^{-4}$ \citep{stubblefield2021variational}, in our model we find only a single GL migrating in phase with the tides. This migration results in a $600$-m wide grounding zone, which is larger than estimated from hydrostatic equilibrium ($2A/\theta=100~\text{m}$). 

To demonstrate tidal flexure, we plot the deviatoric stress component $\tau_{xx}$ at four tidal phases (\autoref{fig:ref_case}a--d). The ice undergoes upward and downward flexure at high and low tides, respectively. At high tide (\autoref{fig:ref_case}a), the stress is concentrated close to the GL, with compression near the top and tension near the bottom, downstream of the GL. This resembles the stress pattern of a beam \citep{timoshenko1955strength}, indicating a region where the ice vertical velocity transitions from the ice-sheet flow to the ice-shelf oscillation with tides. At low tide (\autoref{fig:ref_case}c), the tensile stress dominates the ice-sheet top surface near the GL. The region experiencing tensile stress is larger and located further upstream. At rising tides (\autoref{fig:ref_case}b) and falling tides (\autoref{fig:ref_case}d), $\tau_{xx}$ is tensile at the GL, but the magnitude is smaller than $\tau_{xx}$ at low tide.

The full horizontal tensile stress $\sigma_{xx}$ is considered for hydrofracturing at the lake. Assuming that the lake covers the ice-sheet surface within $|x-\overline{x}_g|\le0.5~\text{km}$, where $\overline{x}_g$ is the time-averaged GL position in a tidal period, we calculate the maximum $\sigma_{xx}$ on the ice-sheet surface within the lake region for any given time, which is denoted $\sigma_{xx,max}$. The temporal variation of $\sigma_{xx,max}$ is shown in \autoref{fig:ref_case}e. In each tidal period, $\sigma_{xx,max}$ reaches its peak at low tide, corresponding to the downward flexural stress in \autoref{fig:ref_case}c. 


The reference case gives the tidal stress at tidal amplitude $A=1$ m. We further consider cases with a series of tidal amplitude from $0$ to $1$ m, and thus obtain a stress--amplitude relationship for sinusoidal semi-diurnal tides, which is referred to as the “$\sigma$--$A$ relationship” from this point forward. However, with solar tides, tidal amplitude is modulated in a two-week cycle. Given the viscoelastic rheology with its history-dependence, such an amplitude modulation might complicate the $\sigma$--$A$ relationship from monochromatic tides. 
\begin{figure}[t]
    \includegraphics[width=16cm]{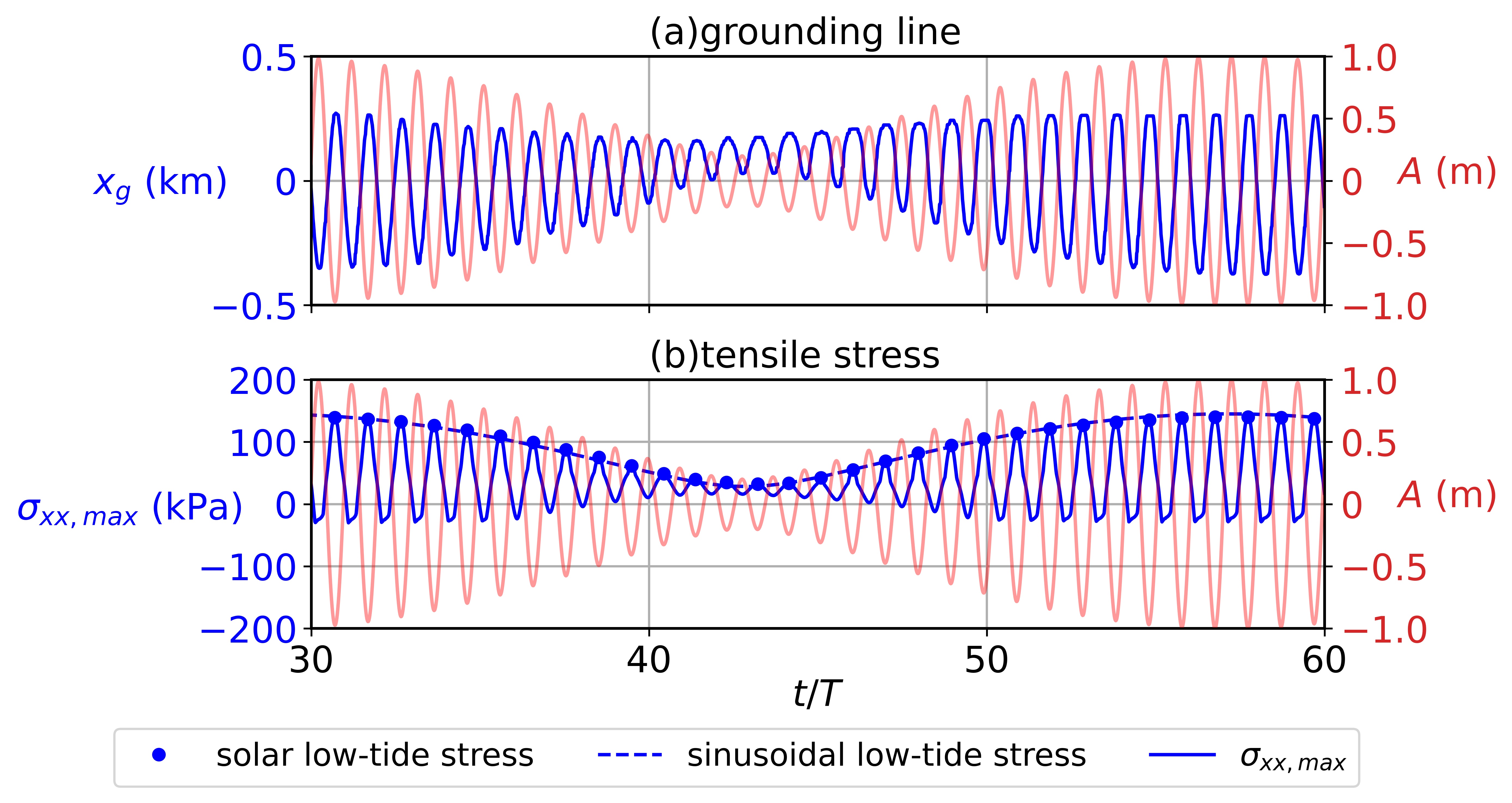}
    \caption{\textbf{(a)} Modulated tidal amplitude (red) and corresponding GL migration (blue). The horizontal axis is time scaled by the tidal period. \textbf{(b)} (blue) Maximum deviatoric tensile stress $\sigma_{xx,max}$ on the ice-sheet surface within the lake region with modulated tidal amplitude. The dots denote the low-tide stress in one tidal period. The dashed blue line is the estimated low-tide stress calculated by the $\sigma$--$A$ relationship from sinusoidal semi-diurnal tides in \autoref{sec:ref_case_stress}.}
    \label{fig:tidal_modulation}
\end{figure}

To explore the $\sigma$--$A$ relationship with solar tides, we replace the sinusoidal tides in Eq. \eqref{eq:tide} with a modulated sine wave over a $14$-day period, with sea-level variation shown in \autoref{fig:tidal_modulation}. Applying this forcing to the reference case, the GL migrates in phase with tides (\autoref{fig:tidal_modulation}a). In each tidal period the low-tide $\sigma_{xx,max}$ tracks the $\sigma$--$A$ relationship for sinusoidal tides (\autoref{fig:tidal_modulation}b), indicating that solar tidal amplitude modulation does not change the $\sigma$--$A$ relationship. Therefore, daily maximum tidal amplitude proves to be a good metric to estimate the daily maximum tidal stress that contributes to hydrofracturing.

\subsection{Linear Elastic Fracture Mechanics model of the hydrofracture}\label{subsec:LEFM}
\begin{figure}[t]
    \includegraphics[width=12cm]{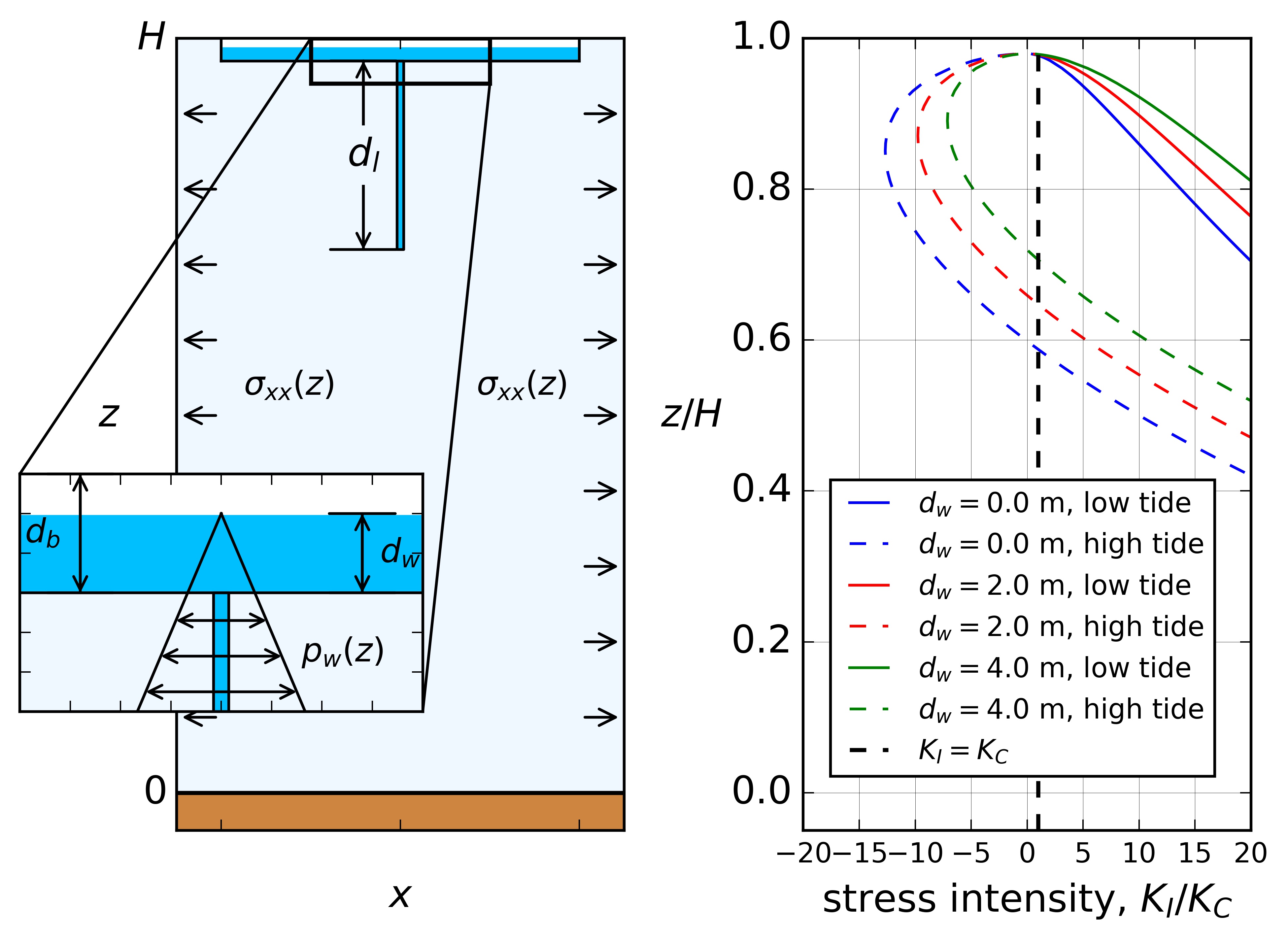}
    \caption{\textbf{(a)} The LEFM model of the hydrofracture. The lake basin with depth $d_b$, is filled with water to a depth $d_w$. Here $d_w$ serves as a measurement of the water pressure $p_w$, as shown in the zoom-in window. Promoted by the tidal stress $\sigma_{xx}\left(z\right)$ and lake-water pressure $p_w$, a vertical fracture with length $d_l$ is initiated from the lake bottom.  \textbf{(b)} A reference case showing $K_{I}/K_{C}$ varying with depth (scaled by the ice thickness) and tidal phases, with $A=1.0$~m, $d_b=10$~m, and $d_{l,init}=0.1$~m. The solid lines represent $K_{I}/K_{C}$ at low tides. The dashed lines represent $K_{I}/K_{C}$ at high tides when upward flexure causes compression, and thus a negative tidal contribution to hydrofracture.}
    \label{fig:LEFM}
\end{figure}
Since hydrofracturing typically occurs on a short timescale over which ice behaves elastically, we consider the fracture propagation in the LEFM framework. The hydrofracture is assumed to be a quasi-static elastic fracture occurring at the location with $\sigma_{xx,max}$ at low tides. The stress that drives its propagation is the sum of the water pressure and tidal stress. The water pressure in the fracture $p_w$ is assumed to be hydrostatic; the tidal stress $\sigma_{xx,max}$ is calculated by the viscoelastic model mentioned above. We use the weight-function method to calculate the stress intensity factor $K_{I}$ \citep{tada2000analysis}. Since at low tides the GL goes downstream and leaves the ice beneath the lake attached to the bedrock, we use a weight function that is designed for ice grounded on rigid bedrock, as suggested by \cite{jimenez2018evaluation}. 

Figure \ref{fig:LEFM}a shows a schematic of the fracture model. The lake basin has a depth $d_b=10$~m (obtained from the DEM) and is filled with water to a depth $d_w$. The horizontal stress $\sigma_{xx}(z)$ represents the low-tide tidal stress, and is obtained from the numerical results with a given tidal amplitude $A$. For a vertical fracture, we can calculate the stress intensity factor $K_{I}$ as a function of its length $d_l$. If $K_{I}$ exceeds the ice fracture toughness $K_{C}$, the fracture can propagate, until $K_{I}=K_{C}$. We assume that lake drainage occurs when a vertical hydrofracture reaches the ice-sheet bottom. 


Note that for the initial fracture, $K_{I}$ is sensitive to its length $d_{l,init}$. However, there is little observational constraint on the lengths of pre-existing fractures at the lake basin. While the choice of $d_{l,init}$ requires further study, the relative importance of the lake pressure versus the tidal amplitude is independent of $d_{l,init}$, which is shown in the model-based criterion below. Here we will choose $d_{l,init}$ such that the model criterion best fits the drainage data.

In \autoref{fig:LEFM}b, assuming that $A=1$~m, we plot a reference case showing $K_{I}$ versus the lake depth $d_l$, under different combinations of $d_w$ ($d_w=0,~2,~4$~m) and tidal phases (low tide and high tide). The vertical dashed line represents the ice fracture toughness $K_{C}$. At low tide, downward flexure gives positive $K_{I}$ near the upper surface. Conversely, at high tides, compression gives negative $K_{I}$ there. Figure \ref{fig:LEFM}b can be used to predict lake drainage: at low tides, for a pre-existing fracture with length $d_{l,init}$, if $K_{I}> K_{C}$ holds for any depth that the fracture can reach, then $A$ and $d_w$ are predicted to induce lake drainage. Repeating this treatment for different combinations of $A$ and $d_w$ yields the model-based drainage criterion, which is shown in \autoref{subsec:drainage_criterion}.

\subsection{Drainage criteria in terms of tidal amplitude and lake depth}
\label{subsec:drainage_criterion}
\begin{figure*}[t]
    \includegraphics[width=14cm]{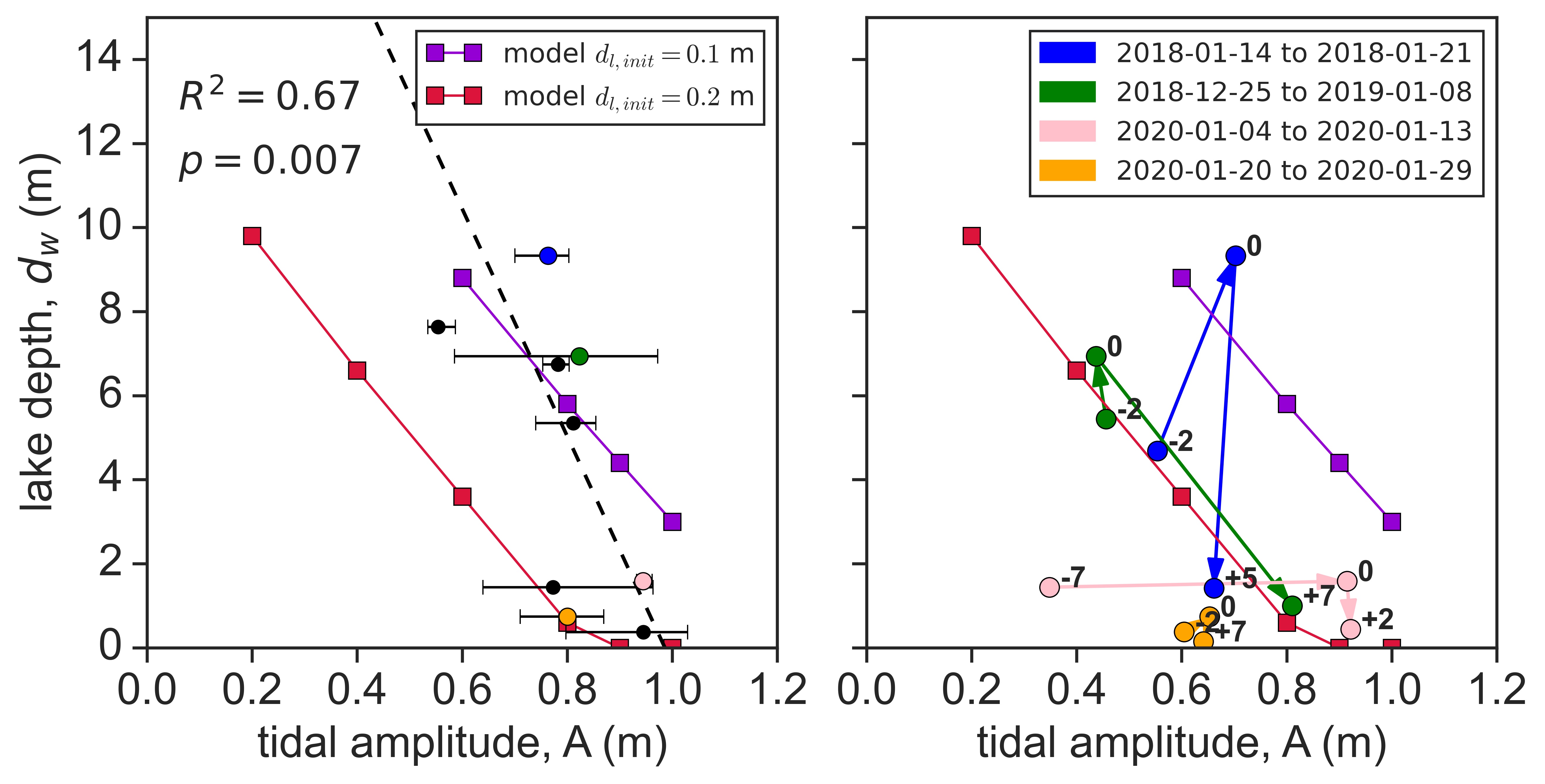}
    \caption{\textbf{(a)} A comparison between the model criterion and the drainage data. Each circle represents one drainage event from \cite{trusel2022repeated}. The horizontal coordinate is the time-averaged daily maximum tidal amplitude during the drainage, with an error bar representing the range of the daily maximum tidal amplitude. The vertical coordinate is the pre-drainage lake depth. The dashed black line is a weighted linear regression of the observations. The red and violet lines are model-based criteria with different initial crack lengths, with squares representing the numerical cases. The four colored circles represent drainage events with best-constrained temporal evolution of lake depth and tidal amplitude. \textbf{(b)} Temporal evolution of coloured events in panel (a). The points labelled “0” represent the day of the drainage. The negative and positive values represent the days before and after the drainage, respectively.}
    \label{fig:criterion}
\end{figure*}

As shown above, at low tides for an initial fracture with a given length $d_{l,init}=0.1$~m, the prediction of hydrofracturing is controlled by two factors: tidal stress and water pressure, measured by tidal amplitude $A$ and lake depth $d_w$. When the combined effect of tidal flexure and lake-water pressure can overcome the ice fracture toughness and ice-overburden stress, the fracture propagates, reaches the ice--bedrock interface, and drains the lake.

We construct a model-based drainage criterion in terms of $A$ and $d_w$ in a 2-D parameter space. The criterion is defined as the marginal conditions at which the initial fracture can reach the ice--bedrock interface. In \autoref{fig:criterion}a, we show two criteria with different initial crack lengths $d_{l,init}$ of $0.1$ and $0.2$~m. When $\left(A,d_w\right)$ of a lake crosses the criterion from bottom-left to top-right, the total tensile stress is large enough that the hydrofracture can reach the bottom of the ice sheet. As shown in \autoref{fig:criterion}a, the increasing tidal amplitude can reduce the lake depth required for hydrofracturing. This provides a measure of the relative importance of tidal flexure to water pressure for lake drainage. Meanwhile, there exists a maximum tidal amplitude above which fracturing can be induced entirely by tidal stress, without a water supply. Supraglacial lakes would not be able to form under such large tidal stress.

In \autoref{fig:criterion}a, we compare the model-based criterion with observations from \cite{trusel2022repeated}. Each circle represents a drainage event. To leading order, the model agrees with the data. The data cluster close to the model-based criterion for realistic values of $d_{l,init}$. A weighted linear regression of the observations also suggests that higher tidal amplitude reduces the lake depth required for drainage. However, the steeper slope of the regression line suggests the dependence of drainage on tidal amplitude is stronger than predicted by our model.

To better demonstrate the drainage process, in \autoref{fig:criterion}b we show four events with the best observational constraint on temporal evolution of lake depth and tidal amplitude. We show measurements from before, during and after the drainage. Note that we simply assume the drainage occurs at the highest water level observed, which is the minimum pre-drainage water level due to the time interval between satellite images. The two events dated 2018-12-25 and 2020-01-04 cross the criterion with $d_{l,init}=0.2$~m in the observational interval of several days. The event dated 2018-01-04 crosses both criteria. The post-drainage states are below the criterion with $d_{l,init}=0.1$~m, representing the end of drainage due to insufficient water supply. Therefore, we believe our model captures the essential physics that controls hydrofracturing in the grounding zone.

\section{Discussion}\label{sec:discussion}
We use a viscoelastic model to predict GL migration and tidal stress of a marine ice sheet. By considering tidally-induced hydrofracturing in an LEFM model, we construct a model-based drainage criterion that accounts for the tidal amplitude and lake depth required for drainage. The criterion agrees with observations to leading order, thus supporting the hypothesis from \cite{trusel2022repeated} that “\textit{hydrofracture is assisted by tidally forced ice flexure inherent to the ice shelf grounding zone}.” Furthermore, the criterion suggests that, in grounding zones, ocean tides might set a maximum volume for supraglacial lakes. We hypothesise that lakes tend to remain at a smaller volume than this critical value due to regular drainage events. This hypothesis could be tested by a statistical study of more supraglacial lake drainage events with local ocean tides.

The tidal-flexure model and drainage criterion can be applied to other marine ice sheets with different rheological properties and bed topography. Here we perform a sensitivity analysis to explore the dependence of GL migration and tidal stress on ice rheology and bedslope angle. From this point froward, we use $\sigma_{xx,max}$ to denote the low-tide maximum tensile stress within the lake region, which is assumed to directly contribute to hydrofracturing. Finally, we discuss the limitations of our model by exploring missing factors that might affect GL dynamics and hydrofracture.

\subsection{Sensitivity to ice Maxwell time}
Ice Maxwell time $\tau$ controls the viscoelastic tidal response of a marine ice sheet. Depending on the Deborah number $De$, the ratio of $\tau$ to the tidal period $T$, the tidal response can be dominated by either elasticity ($De\ll1$) or viscosity ($De\gg1$). Here, we explore the tidal response with a varying $\tau$, representing a transition from viscoelastic to viscous rheology.
\begin{figure*}[t]
   \centering{\includegraphics[width=14cm]{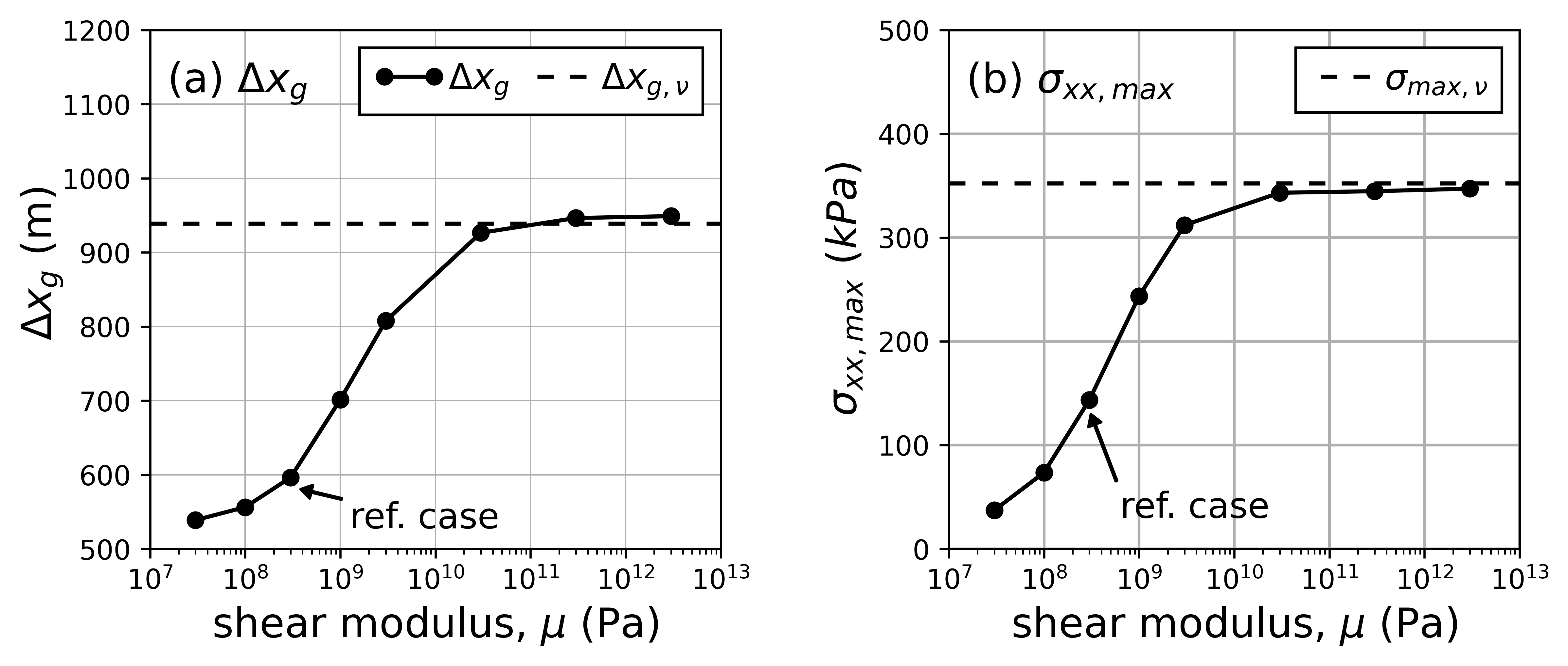}}
    \caption{\textbf{(a)} The grounding-zone width $\Delta x_g$ (solid line), defined as $\Delta x_g = \max{\{x_r\}}-\min\{x_l\}$ as a function of shear modulus $\mu=3\times10^{7}$ to $3\times10^{12}$ Pa, with $x_l$ and $x_r$ denote the left and right GL, respectively. The dashed line shows $\Delta x_{g,\nu}$, the grounding-zone width in the viscous limit ($\mu\rightarrow\infty$). \textbf{(b)} Maximum tensile stress $\sigma_{xx,max}$ versus $\mu$. The numerical reference case in \autoref{sec:results} is labelled.}
    \label{fig:maxwell_time}
\end{figure*}

With the viscosity governed by Glen's Flow law, we consider cases with varying shear modulus $\mu$, thus varying Maxwell time. All other parameters are set to their reference values. We define a grounding-zone width, $\Delta x_g$, as the range of GL in a tidal period. In \autoref{fig:maxwell_time}a, $\Delta x_g$ is plotted versus $\mu$. With the bedslope $\theta=2\times10^{-2}$, only one GL is found to migrate with the tides, giving a grounding-zone width of about $0.5-1$ km. As $\mu\rightarrow\infty$, the tidal response becomes purely viscous and the grounding zone widens and reaches its viscous limit $\Delta x_{g,\nu}$. In \autoref{fig:maxwell_time}b, $\sigma_{xx,max}$ increases with increasing $\mu$ and converges to the viscous stress calculated by the viscous model \citep{stubblefield2021variational}.

Ice Maxwell time is crucial in determining GL migration and tidal stress, because viscous rheology tends to increase the width of a grounding zone and produce larger tensile stress than viscoelastic rheology. Therefore, our model indicates the importance of using viscoelastic rheology with an accurate ice Maxwell time to predict the magnitude of tidal stress. Given the dependence of $\Delta x_g$ on $\mu$ (\autoref{fig:maxwell_time}a), it may be possible to infer ice mechanical properties from observations on the range of GL migration.

\subsection{Sensitivity to bedslope angle}
\begin{figure*}[t]
    \includegraphics[width=14cm]{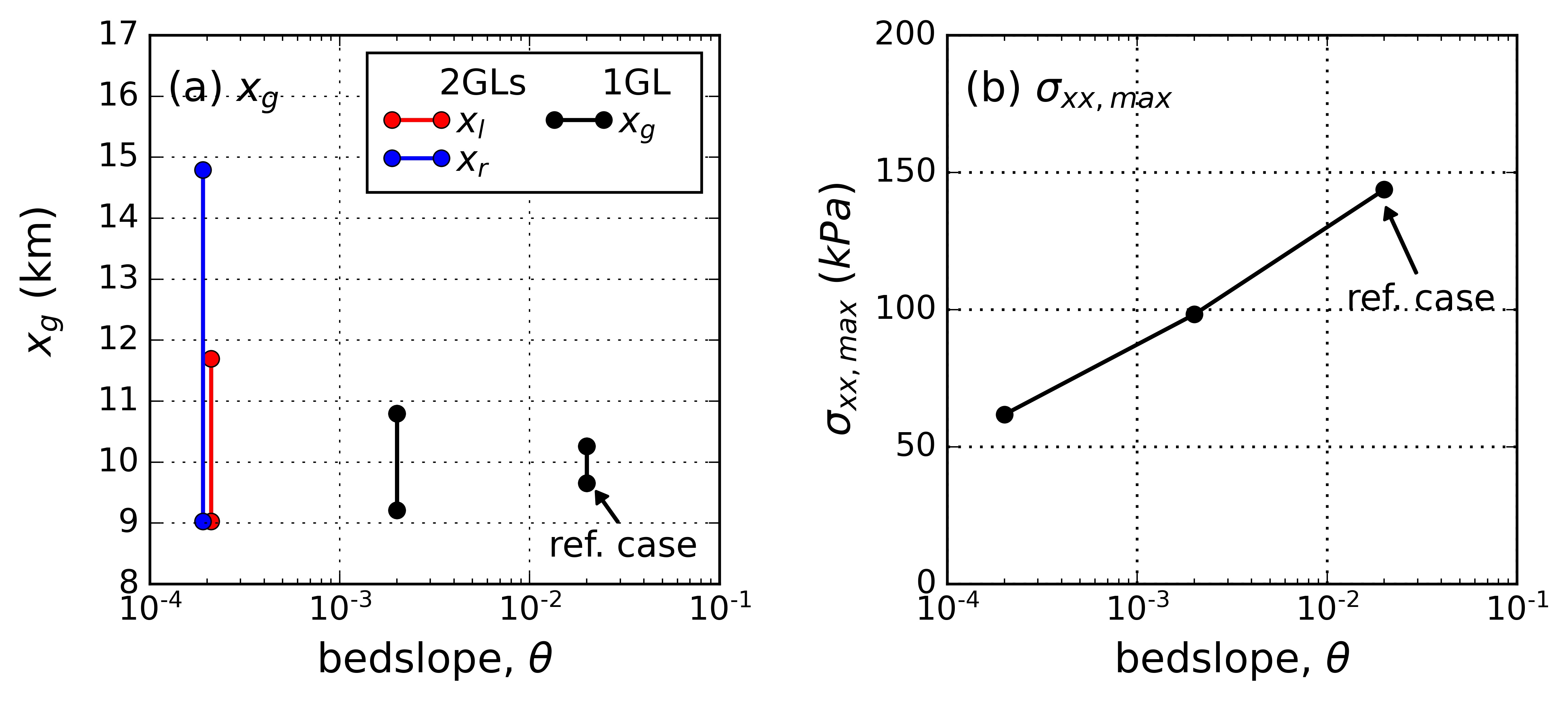}
    \caption{\textbf{(a)} The range of the GL position in one tidal period as a function of bedslope angle $\theta=2\times10^{-4}$, $2\times10^{-3}$, $2\times10^{-2}$. When $\theta=2\times10^{-4}$, there are two GLs. The left and right GLs are denoted $x_l$ and $x_r$, respectively. The other two cases give single GL $x_g$ shown by the black line. \textbf{(b)} Maximum tidal stress $\sigma_{xx,max}$ versus $\theta$.}
    \label{fig:maxwell_time_2}
\end{figure*}

The above discussion shows how tidal response varies with shear modulus, with a characteristic bedslope $\theta=2\times10^{-2}$ at the Amery Ice Shelf GL. Here we extend the results to different bedslopes and explore how the tidal response of a GL would change with local bathymetry. We consider three marine ice sheets with bedslope $\theta=2\times 10^{-4}$, $2\times 10^{-3}$, and $2\times 10^{-2}$, with all other parameters set to be the same as the reference case. 
For simplification, we focus on the effect of $\theta$ and keep the basal friction coefficient $C$ in Eq.\eqref{eq:basal_friction} fixed. Because of this, the modelled surface velocity varies away from the observed value $u_0$, but maintains the same order of magnitude. The GL migration is shown in \autoref{fig:maxwell_time_2}a. Different from the single GL shown above, the low-bedslope regime $\theta=2\times 10^{-4}$ is characterized by double GLs at low tides. Between the left GL at $x_l$ and the right GL at $x_r$, the ice sheet is lifted due to a water layer trapped underneath \citep{stubblefield2021variational}, forming a “low-tide grounding zone.” For the other two cases, only a single GL is found, with the range of the GL decreasing with $\theta$. Moreover, the maximum tidal stress monotonically increases with $\theta$ (\autoref{fig:maxwell_time_2}b). For a specific grounding line, the local basal topography and characteristic bedslope angle can be constrained by observations \citep{Fretwell2013bedmap2}. Thus, the uncertainties of the modelled tidal GL migration and stress mainly come from the rheological model.

Our sensitivity analysis is based on the assumption of a fixed basal friction parameter. To estimate the tidal stress at a specific geographical location, the tidal response should be modelled with a basal sliding law that matches observed surface velocities. Then, a drainage criterion can be set up to estimate the local tidal contribution to supraglacial lake drainage.

\subsection{Limitations}
Deviations between the model-based criterion and data regression indicate that we may be underestimating the tidal contribution to hydrofracturing (\autoref{fig:criterion}). Currently, we assume a stress-free top surface and constant ice properties in the viscoelastic flow model. However, the supraglacial lake can induce additional stress in the surrounding ice, particularly on floating portions of the grounding zone that allow downward flexure \citep{macayeal2013flexural}. Meanwhile, the existence of ice fatigue due to stress oscillations can weaken ice strength and promote hydrofracturing \citep{borstad2012damage,lhermitte2020damage}. A better approach may be to directly consider the supraglacial lake and the ice damage within the 2-D viscoelastic model.

Another limitation arises from our assumption of hydrostatic water pressure on the ice--ocean interface. The pressure gradient induced by tidally modulated subglacial water flow can cause elastic flexure in ice sheets close to the grounding line \citep{warburton2020tidal}. Furthermore, ocean tides can change the effective pressure at the bed and lubricate the ice--bedrock interface, leading to a variation of basal friction that is not accounted for by the sliding law in our model \citep{gudmundsson2011ice}. Thus, it is important to incorporate the subglacial hydrology in simulating the tidal response of a marine ice sheet.

Limitations also come from the data availability. Relative to the tidal period and lake-drainage period, the lower temporal resolution of the remotely sensed observations might obscure the true lake depth and tidal amplitude at the time of drainage. Field measurements and satellite images of supraglacial lake drainage with a higher observational frequency could improve our understanding of tidally-induced drainage.

\conclusions  
\label{sec:conclusion}
Our study of tidally-induced stress and hydrofracture propagation in a viscoelastic marine ice sheet grounding zone supports the hypothesis on tidally-induced supraglacial lake drainage proposed by \citet{trusel2022repeated}. We further propose a model-based criterion for lake drainage that indicates how ocean tides and lake depth together determine whether supraglacial lakes will drain via hydrofracture at the Amery Ice Shelf GL. Importantly, the criterion indicates that grounding-zone lakes tend to remain at a smaller volume than a critical value controlled by tidal flexure. For similar, tidally-modulated marine ice sheets, our results suggest that ocean tides can generate significant stress near the grounding line, potentially increasing the vulnerability of ice sheets to hydrofracturing in grounding zones where lakes form. Our work serves as an initial attempt to analyse the tidal effect on hydrofracturing, which helps to better explain the role of ocean tides in driving supraglacial lake drainage, calving events, and grounding line migration in marine ice sheets. 


\codedataavailability{The code and data used for the reference case in \autoref{sec:results} is available at the online GitHub repository \linebreak\url{https://github.com/HwenZhang/TidalHydroFrac.git}. It can be modified to reproduce the results presented in \autoref{sec:discussion} using the values of parameters provided in the text.} 


\appendix
\section{Convergence test}    
\label{apdx:convergence}
\begin{figure}[t]
    \includegraphics[width=14cm]{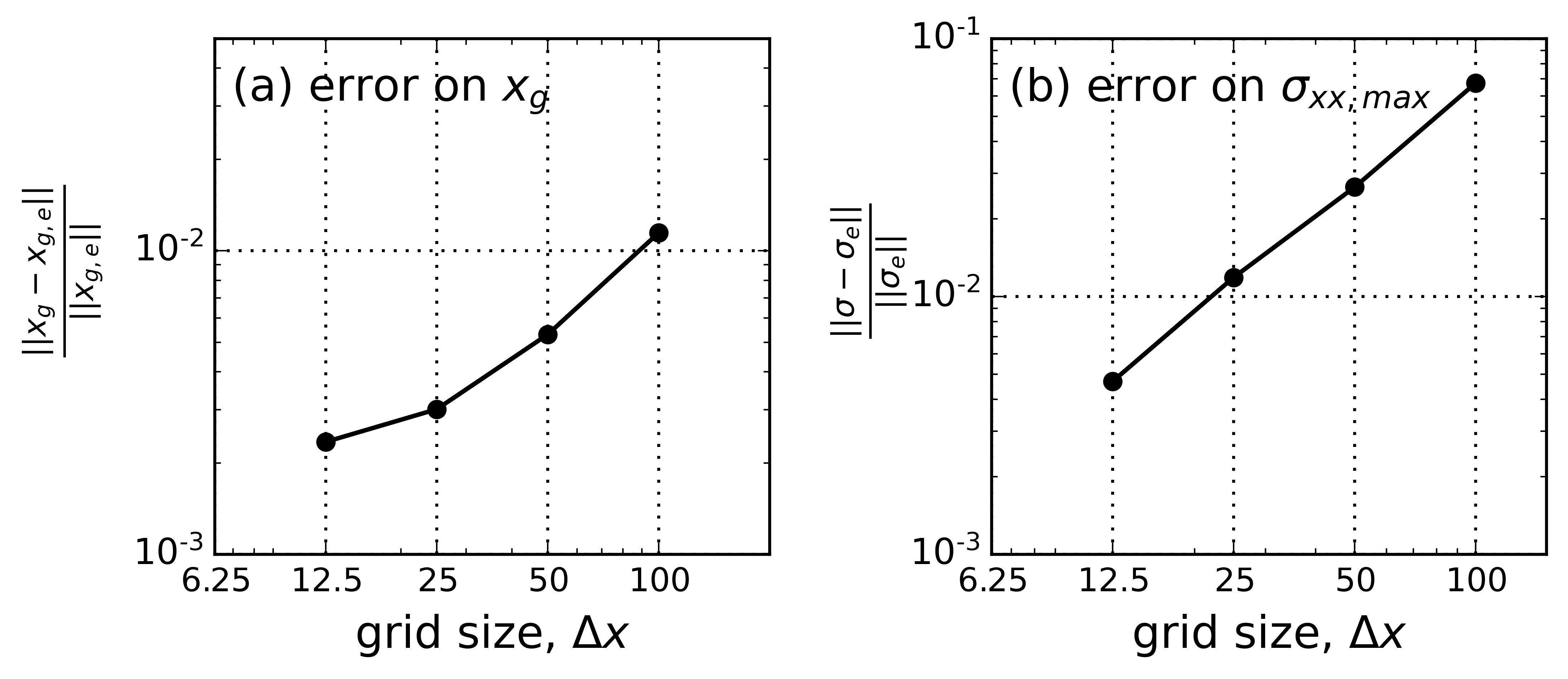}
    \caption{Convergence of (a) GL position and (b) maximum tensile stress $\sigma_{xx,max}$ with decreasing element size $\Delta x$ ($12.5$~m, $25$~m, $50$~m, $100$~m). For simplicity, we denote $\sigma_{xx,max}$ by $\sigma$ without causing any confusion. Here ${{x}_{g,e}}$ and ${\sigma_{e}}$ denote the exact solution to the GL position and maximum tensile stress $\sigma_{xx,max}$, respectively. $||\cdot||$ is the $L^2$ norm.}
    \label{fig:convergence}
\end{figure}

The convergence test shows the results are mesh-independent. Considering a marine ice sheet with bedslope $\theta=10^{-3}$ and friction coefficient $C=7\times10^{5}$, we use the fine-grid solution ${{x}_{g,e}}$, ${{\sigma}_{e}}$ ($\Delta x = 6.25$m) as the exact solution. Here ${{x}_{g,e}}$ denotes the time series of the exact GL position, and ${\sigma_{e}}$ denotes the time series of the exact maximum tensile stress on the ice-sheet surface within the lake region $|x-\bar{x}_g|\le 0.5~\text{km}$. As $\Delta x$ decreases, the GL position $x_g$ and maximum tensile stress $\sigma_{xx,max}$ linearly converge to the fine-grid solution (\autoref{fig:convergence}).

\appendixfigures  

\section{Simulation with real bed topography}\label{apdx:real_bed}
\begin{figure*}[t]
    \includegraphics[width=16cm]{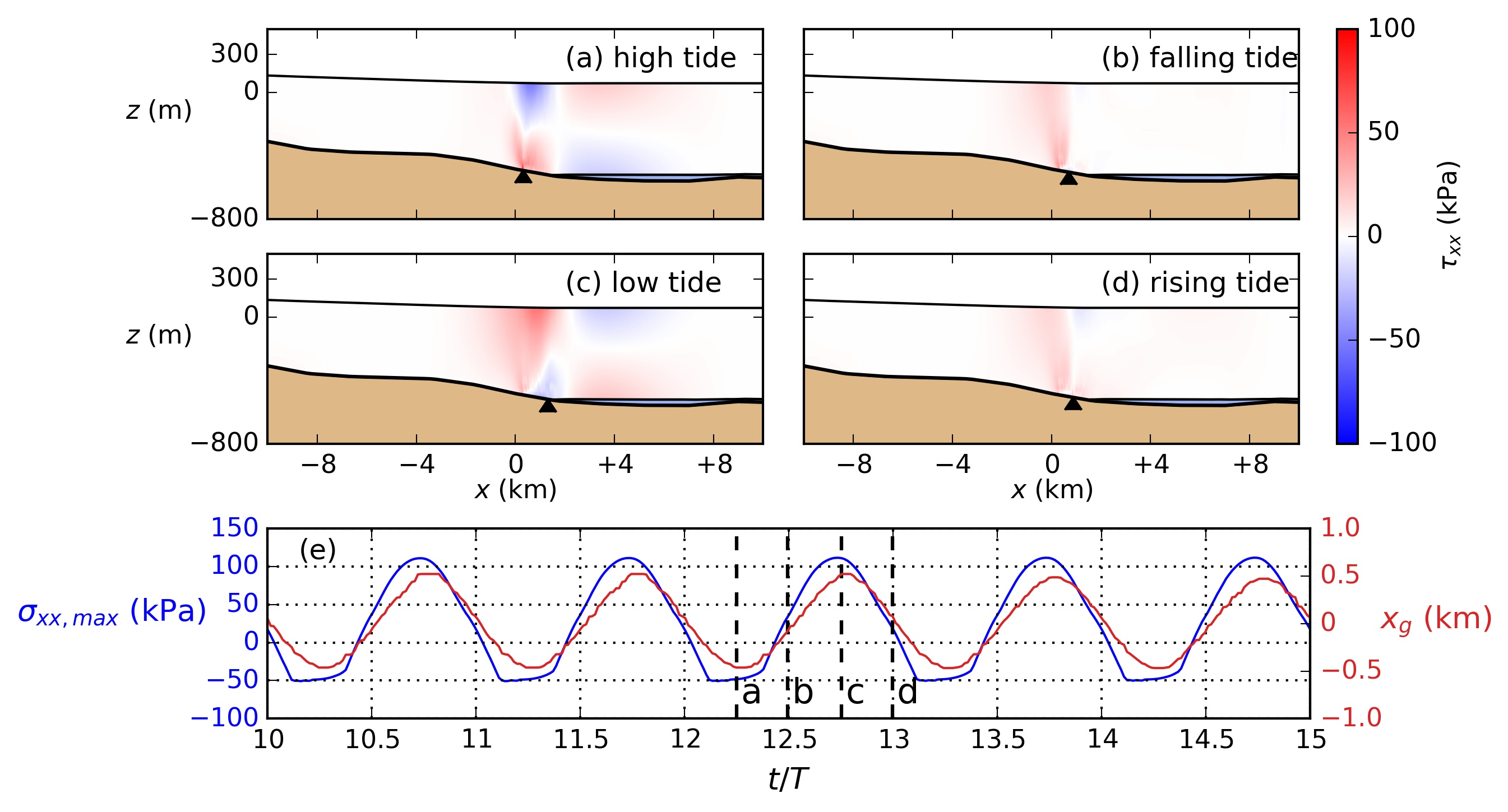}
    \caption{Tidal response of the Amery Ice Shelf with real bed topography. \textbf{(a)--(d)} Deviatoric tensile stress $\tau_{xx}$ in one tidal period. \textbf{(e)} The maximum tensile stress $\sigma_{xx,max}$ (blue) on the top boundary within the lake region ($\bar{x_g}-0.5~\text{km}\le x \le \bar{x_g}+ 0.5~\text{km}$) and the GL position $x_g$ (red) versus time (scaled by the tidal period $T$) with positive values representing downstream migration. Vertical dashed lines show the time of panels (a)-(d). }
    \label{fig:real_bed_case}
\end{figure*}

In \autoref{fig:real_bed_case} we present model results using the real bed topography shown in \autoref{fig:amery}e \citep{morlighem2017bedmachine, morlighem2020deep}. For comparison, the physical properties of ice and basal slipperiness $C$ are kept the same as in idealised models with a linear bed. The jagged variation of $x_g(t)$ is a consequence of the use of coarse grids near the grounding line for convergence. While the tidal stress and grounding-zone width are modified by bed undulation, the results have an equivalent order of magnitude to the case with linear bed topography, indicating the importance of tidal stress regardless of bed roughness. Therefore, we use the linear bed topography in the model.

\noappendix       




\appendixfigures  

\appendixtables   


\authorcontribution{\textbf{Hanwen Zhang}: Conceptualization, Methodology, Software, Validation, Formal analysis, Investigation, Data Curation, Writing - Original Draft, Writing - Review \& Editing, Visualization. \textbf{Richard Katz}: Conceptualization, Software, Resources, Writing - Review \& Editing, Supervision, Project administration, Funding acquisition. \textbf{Laura Stevens}:  Conceptualization, Resources, Writing - Review \& Editing, Supervision, Project administration.} 

\competinginterests{The authors declare that they have no conflict of interest.} 


\begin{acknowledgements}
The authors thank L.~D.~Trusel and A.~Fatula for help in interpretation of their lake-drainage observations, and I.~Hewitt, C.-Y.~Lai, and the RIFT-O-MAT group for discussions on grounding line dynamics and model set-up. This research received funding from the European Research Council under Horizon 2020 research and innovation program grant agreement number 772255.
\end{acknowledgements}







\bibliographystyle{copernicus}
\bibliography{cas-refs}

\end{document}